\documentclass[11pt]{article}
\usepackage{moriond,epsfig}

\def\be{\begin{equation}}
\def\ee{\end{equation}}
\def\bea{\begin{eqnarray}}
\def\eea{\end{eqnarray}}

\begin{document}
\vspace*{4cm}
\title{Study light scalar meson property from heavy meson decays}

\author{ Cai-Dian L\"u$^a$ and Wei Wang$^b$}

\address{$^a$Institute of High Energy Physics,  Chinese Academy
 of Sciences, Beijing 100049, PR China\\
 $^b$Istituto Nazionale di Fisica Nucleare, Sezione di Bari, Bari 70126, Italy}

\maketitle\abstracts{ In the SU(3) symmetry limit, the ratio
$R\equiv\frac{{\cal B}(D^+\to f_0l^+\nu)+
 {\cal B}(D^+\to \sigma l^+\nu)}{{\cal B}(D^+\to
 a_0^0l^+\nu)}$ is equal to 1 if the scalar mesons are
  $\bar qq$ states, while it is 3 if these mesons are
  tentraquark states. This ratio provides a model-independent way to
  distinguish the
descriptions for light scalar mesons . It also applies to the
$B^-\to Sl^-\bar\nu$ and $\bar B^0\to J/\psi(\eta_c) S$ decays. The
SU(3) symmetry breaking effect is found to be under control, which
will not spoil our method. The branching fractions of the $D^+\to
Sl^+\nu$, $B^-\to Sl^-\bar\nu$ and $\bar B^0\to J/\psi(\eta_c) S$
decays  roughly have the order $10^{-4}$, $10^{-5}$ and $10^{-6}$,
respectively. The
 B  factory experiments and ongoing BEPC-II  experiments are able to measure
these channels and accordingly to provide the detailed information
of the scalar meson inner structure.}


In spite of the striking success of QCD theory for strong
interaction, the underlying structure of the light scalar mesons is
still under
controversy.\cite{Spanier:2008zz,Godfrey:1998pd,Close:2002zu} To
understand the internal structure of scalar mesons is one of the
most interesting topics in hadron physics for several decades.
Irrespective of the dispute on the existence of $\sigma$ and
$\kappa$ mesons, scalar mesons have been identified as ordinary
$\bar qq$ states, four-quark states or meson-meson bound states or
even those supplemented with a scalar glueball. Due to the unknown
nonperturbative properties of QCD, there is almost no
model-independent way to effectively solve these old puzzles.

Most of the studies up to now, concentrated on the decay property of
scalar mesons. It is interesting to study the production property of
the scalar mesons, especially the production from heavy meson
decays. At present, there are many experimental studies on the
production of scalar mesons in nonleptonic $D$ decays. For example,
branching ratios of $D^+\to \sigma\pi^+$ and $D^+\to f_0\pi^+$ have
the order of $10^{-3}$ and $10^{-4}$,
respectively.\cite{Amsler:2008zzb} On the theoretical side compared
with nonleptonic $D$ decays, semileptonic $D^+(B^+)\to S l^+ \nu$
decays only contain one scalar meson in the final state, where the
heavy quark effective theory can be used. This could be the better
candidate to probe different structure scenarios of scalar mesons.

In this work, we will only focus  on the two-quark and the
four-quark scenarios for scalar mesons.  We propose a
model-independent way to distinguish these two descriptions through
the semileptonic $B^-\to Sl\bar\nu$ and/or $D^+\to Sl^+\nu$ decays,
where $S$ denotes a scalar meson among $a_0(980),f_0(980)$ and
$\sigma$.\cite{wl} These two kinds of decays are clean as they do
not receive much pollution from the strong interactions. In $B$
decays, the lepton pair can also be replaced by a charmonium state
since they share the same properties in the flavor SU(3) space. For
example, the $\bar B^0\to J/\psi(\eta_c) S$ decays  are probably
much easier for the experiments to observe.


A number of scalar mesons have been discovered experimentally. Among
them, there are 9 mesons below or near 1 GeV, which form  a nonet
   consisting of
$\sigma,\kappa,f_0(980)$ and $a_0(980)$. Hereafter we will use $f_0$
and $a_0$ to abbreviate $f_0(980)$ and $a_0(980)$ for simplicity. In
the $\bar qq$ picture, scalar mesons are viewed as P-wave
states,\cite{Tornqvist:1995kr} whose flavor wave functions are given
by
\begin{eqnarray}
 &  |\sigma\rangle =\frac{1}{\sqrt 2} (|\bar uu\rangle +|\bar
 dd\rangle)\equiv |\bar nn\rangle,\;\;\; |f_0\rangle =|\bar
 ss\rangle,\;\;\;
  |a_0^0\rangle =\frac{1}{\sqrt 2} (|\bar uu\rangle -|\bar
 dd\rangle),\;\;\; |a_0^-\rangle =|\bar u d \rangle,\;\;\;
 |a_0^+\rangle = |\bar d u \rangle.\nonumber\\
&   |\kappa^-\rangle = |\bar u s \rangle,\;\;\;
 |\bar\kappa^0\rangle = |\bar d s \rangle,\;\;\;
 |\kappa^0\rangle = |\bar s d \rangle,\;\;\;
 |\kappa^+\rangle = |\bar s u \rangle.
\end{eqnarray}
In this picture, $f_0$ is mainly made up of $\bar ss$, which is
supported by the large production rates in $J/\psi \to \phi f_0$ and
$\phi\to f_0\gamma$ decays.\cite{Amsler:2008zzb} Meanwhile, the
experimental data also indicate the nonstrange component of $f_0$:
the branching fraction of $J/\psi \to \omega f_0$ is comparable with
that of $J/\psi\to \phi f_0$. To accommodate with the experimental
data, $f_0$ is supposed to be the mixture of $\bar nn$ and $\bar ss$
as
\begin{eqnarray}
 |f_0\rangle = |\bar ss\rangle \cos\theta +|\bar nn\rangle
 \sin\theta, &
 |\sigma\rangle =-|\bar ss\rangle \sin\theta +|\bar nn\rangle
 \cos\theta.\nonumber
\end{eqnarray}
With various available experimental data, the mixing angle $\theta$
is constrained as $
 25^\circ<\theta <40^\circ,  140^\circ <\theta
 <165^\circ$.\cite{Cheng:2005nb}

The classical $\bar qq$ picture meets with several difficulties. In
particular it is difficult to explain the fact that the strange
meson $\kappa$ is lighter than the isotriplet mesons $a_0$, and the
isosinglet meson $f_0$ has a degenerate mass with $a_0$, since $s$
quark is expected to be heavier than $u/d$ quark. Inspired by these
difficulties, other candidate scenarios are proposed.  In
Ref.~\cite{Jaffe:1976ig}, scalar mesons are identified as
diquark-diquark states.  In the SU(3) flavor space, the two quarks
can form two multiplets as
$ 3\otimes 3 =\bar 3 \oplus 6$,
while the other two antiquarks reside in $3$ or $\bar 6$ multiplets.
The diquark in a scalar meson is taken to be totally antisymmetric
for all quantum numbers, color antitriplet, flavor antitriplet, spin
0. The lightest $q^2(\bar q)^2$ states make a flavor nonet, whose
internal structure   is given as:
\begin{eqnarray}
 &|\sigma\rangle = \bar uu \bar dd,\;\;\; |f_0\rangle =|\bar nn\bar
 ss\rangle,\;\;\;
 |a_0^0\rangle = \frac{1}{\sqrt 2} (\bar uu-\bar dd)\bar ss,\;\;\; |a_0^+\rangle =|\bar d u \bar ss\rangle, \;\;\;
 |a_0^-\rangle =|\bar ud \bar ss\rangle. \nonumber\\
 &|\kappa^+\rangle= |\bar s u\bar dd\rangle,\;\;\;  |\kappa^0\rangle =|\bar s d\bar
 uu\rangle,\;\;\;
 |\bar\kappa^0\rangle = |\bar d s\bar uu\rangle,\;\;\;  |\kappa^-\rangle =|\bar u s\bar
 dd\rangle.
\end{eqnarray}
Taking the mixing into account, the isosinglet mesons are expressed
as
\begin{eqnarray}
 |f_0\rangle= |\bar nn\bar ss \rangle\cos\phi +|\bar uu \bar
 dd\rangle\sin\phi, &
 |\sigma\rangle= -|\bar nn\bar ss \rangle\sin\phi +|\bar uu \bar
 dd\rangle \cos\phi,\nonumber
\end{eqnarray}
where the $\phi$ between $f_0$ and $\sigma$ meson is constrained as$
  \phi=(174.6^{+3.4}_{-3.2})^\circ$.\cite{Maiani:2004uc}


\begin{figure}
\includegraphics[scale=0.32]{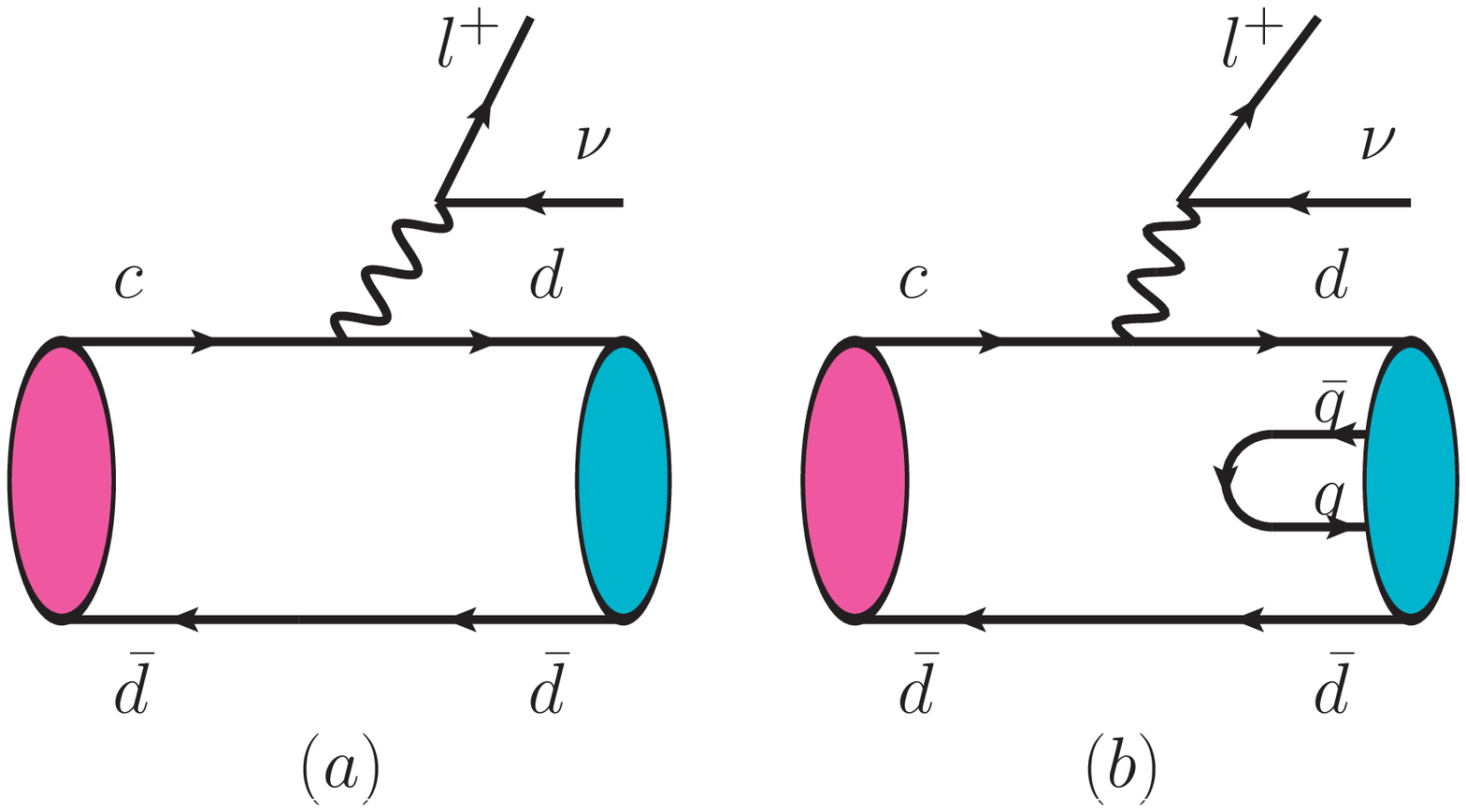}
\hspace{1.cm}
\includegraphics[scale=0.32]{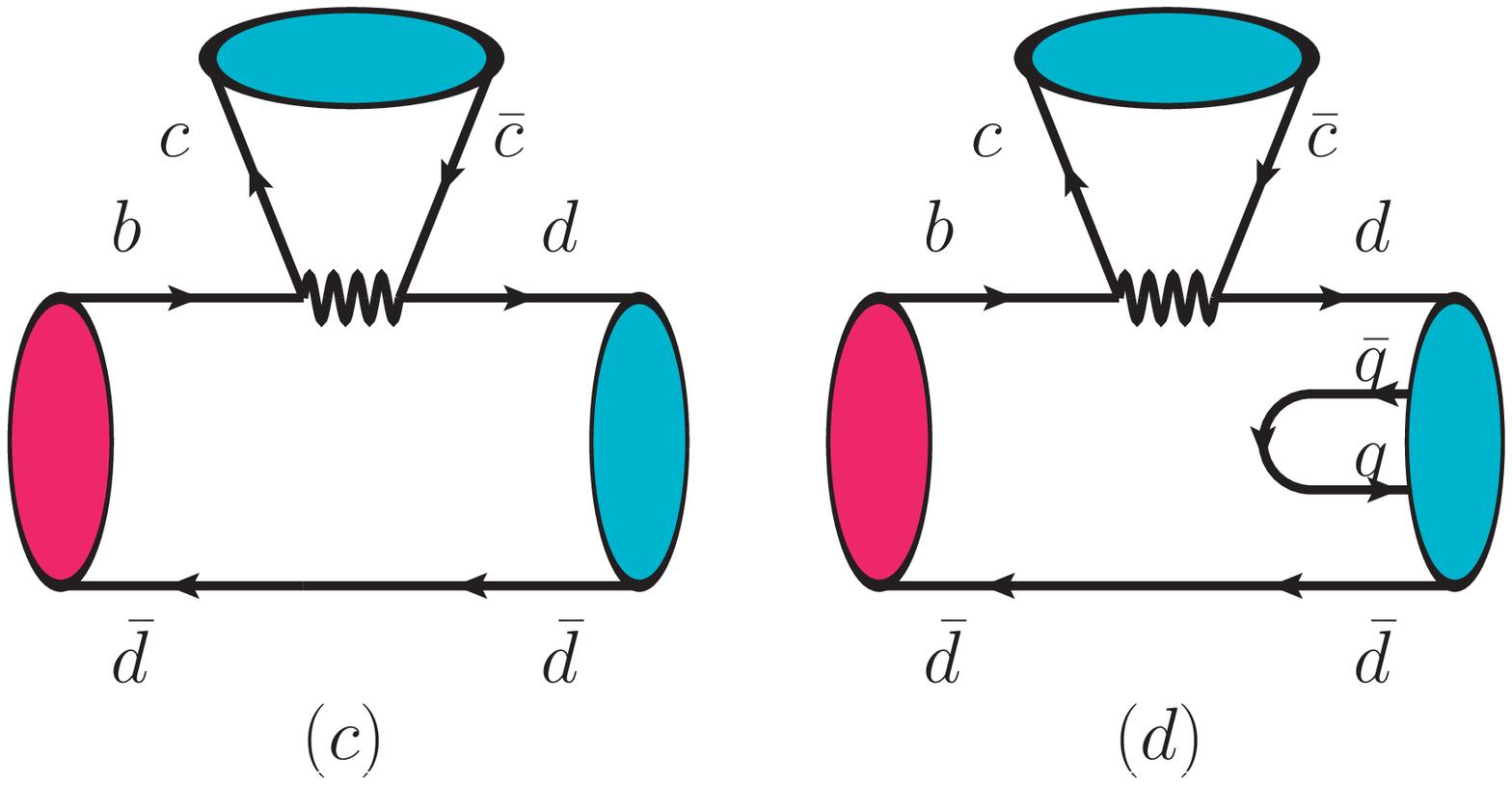}
\caption{Feynman diagrams for $D^+$ decays into a scalar meson and
$B\to J/\psi (\eta_c) S$ decays. The diagrams (a,c) are for
two-quark picture, while the diagrams (b,d) are for the four-quark
mesons. } \label{diagram:Feyn}
\end{figure}

%
%
%

The Feynman diagrams for $D^+\to Sl^+\nu$ decays and $B\to J/\psi
(\eta_c) S$ decays in two different pictures are given in
Fig.~\ref{diagram:Feyn}. The diagrams (a,c) are for the two-quark
scenario, while the diagrams are for the four-quark scenario. If a
scalar meson is made of $\bar qq$, the light quark is generated from
the electroweak vertex and the antiquark  $\bar d$ serves as a
spectator. Thus only the component $\bar d d$ contributes to
semileptonic $D$ decays. In the SU(3) symmetry limit, decay
amplitudes of $D\to f_0(\sigma)l\nu$ channels under the $q\bar q$
picture have the following relation
\begin{eqnarray}
 {\cal A}(D^+\to f_0l^+\nu)=-\sin\theta \hat {\cal A},&
 {\cal A}(D^+\to \sigma l^+\nu)=-\cos\theta\hat {\cal A},
\end{eqnarray}
where the transition amplitude $ \hat {\cal A}$ is defined as $
 \hat {\cal A}\equiv {\cal A}(D^+\to a_0^0l^+\nu).$
This leads to
\begin{eqnarray}
 {\cal B}(D^+\to a_0^0l^+\nu)=
 {\cal B}(D^+\to f_0l^+\nu)+
 {\cal B}(D^+\to \sigma l^+\nu).
 \label{sumrule1}
\end{eqnarray}
One may worry about the accuracy of our results because of the
possible large QCD scattering effect. However if we use the hadron
picture, we can still get the same result. The $d\bar d$ pair
produced from the weak interaction in Fig.\ref{diagram:Feyn}(a) can
form isospin 0 and isospin 1 states with the ratio of 1:1. Although
the scattering can mix between states, the non-perturbative QCD
interactions conserve the isospin. Therefore the sum of production
rates of isospin 0 states on the right hand side of
eq.(\ref{sumrule1}) is always equal to production rates of the
isospin 1 states on the left hand side of eq.(\ref{sumrule1}). The
isospin breaking effect in strong interaction is negligible.

If a scalar meson is composed of four quarks, besides the light
quark from the electroweak vertex and the spectator, another $\bar
qq$ pair is generated from the QCD vacuum. The  decay amplitudes are
given as
\begin{eqnarray}
 {\cal A}(D^+\to f_0l^+\nu)=-(\cos\phi+\sqrt 2\sin\phi)\hat {\cal
 A},&
 {\cal A}(D^+\to \sigma l^+\nu)=(\sin\phi-\sqrt 2\cos\phi)\hat {\cal
 A},
\end{eqnarray}
which gives
\begin{eqnarray}
 {\cal B}(D^+\to a_0^0l^+\nu)=\frac{1}{3}[{\cal B}(D^+\to f_0l^+\nu)+
 {\cal B}(D^+\to \sigma l^+\nu)].\nonumber
\end{eqnarray}
It is meaningful to define the ratio of partial decay widths
\begin{eqnarray}
 R\equiv\frac{{\cal B}(D^+\to f_0l^+\nu)+
 {\cal B}(D^+\to \sigma l^+\nu)}{{\cal B}(D^+\to
 a_0^0l^+\nu)}.\label{rule1}
\end{eqnarray}
Clearly, the ratio is 1 for the two-quark description, while it is 3
for the four-quark description of scalar mesons. Similarly for
semileptonic $B\to Sl\bar\nu$ decays, the charm quark in
Fig.\ref{diagram:Feyn} is replaced by a bottom quark and the $\bar
d$ quark is replaced by a $\bar u$ quark, while leptons are replaced
by their charge conjugates. We have the same sum rules
\begin{eqnarray}
 R&=&\frac{{\cal B}(B^-\to f_0l^-\bar\nu)+
 {\cal B}(B^-\to \sigma l^-\bar\nu)}{{\cal B}(B^-\to
 a_0^0l^-\bar\nu)}
 ~=~\left\{ \begin{array}{cc}1 & \mbox{two~ quark}\\
 3 & \mbox{tetra-quark} \end{array} .\right . \label{rule2}
\end{eqnarray}

The semileptonic $D/B$ decays are clean, which do not receive much
pollution from the strong interaction. But since the neutrino is
identified as missing energy, the efficiency to detect these
channels may be limited.
The lepton pair can also be replaced by some other SU(3) singlet
systems such as a $J/\psi$ or $\eta_c$ meson. Replacing the lepton
pair by the $J/\psi$ and replacing $B^-$ by a $\bar B^0$ state (a
different spectator antiquark will not change the relation) in
Eq.~(\ref{rule2}), one can easily obtain the similar sum rules for
the branching fractions
\begin{eqnarray}
 R&=&\frac{{\cal B}(\bar B^0\to f_0J/\psi(\eta_c))+
 {\cal B}(\bar B^0\to \sigma J/\psi(\eta_c))}{{\cal B}(\bar B^0\to
 a_0^0J/\psi(\eta_c))}
 ~=~\left\{ \begin{array}{cc}1 & \mbox{two~ quark}\\
 3 & \mbox{tetra-quark} \end{array} .\right . \label{rule3}
\end{eqnarray}


Since we   used   SU(3) symmetry to obtain these relations, it is
necessary to estimate the size of SU(3) symmetry breaking effects.
For example, the isospin singlet scalar mesons have different
masses, which can change the phase space in the semileptonic $D/B$
decays. Fortunately, this SU(3) breaking effect can be well studied,
which almost does not depend on the internal structure of scalar
mesons or the strong interactions.  The mass of $f_0$ is well
measured but the mass of $\sigma$ meson has large uncertainties
$m_\sigma=(0.4-1.2)$ GeV. This big range of  masses indeed induces
large differences to $D$ decays, since the $D$ meson mass is only
1.87GeV. The branching ratio of the semileptonic decay is affected
by a factor of $(0.31 -5.4)$ depending on the mass of the $\sigma$
meson. Therefore the sum rule in eq.(\ref{rule1}) is not good unless
the $\sigma$ meson mass is well measured.
 But in $B$ meson decays,   the
sum rule in eq.(\ref{rule1})  will not be affected sizably, since
the $\sigma$ meson mass is negligible compared with the large $B$
meson mass.
 Numerically,
this correction factor in $B$ decays is $ (0.9-1.1)$.

Another SU(3) breaking effect comes from the decay form factors of
various scalar mesons.  In the two-quark scenario, only the $\bar
dd$ component contributes to the form factors shown in the  diagram
(a) of Fig.\ref{diagram:Feyn}.  The  SU(3) symmetry breaking effect
to the form factors is thus negligible. In the four-quark scenario,
there are $\bar uu\bar dd$ component in $f_0$ and $\sigma$ meson
state, which is different from the internal structure of $a_0$ (with
a pair of $\bar ss$). From the  diagram (b) of
Fig.\ref{diagram:Feyn}, one can see that it would be easier to
produce the $\bar uu$ quark pair from the vacuum than the $\bar ss$
quark pair since the $\bar uu$ quark pair is lighter. The SU(3)
symmetry breaking effects may make the form factor of $D/B \to
\sigma$ and $D/B \to f_0$ larger than that of $D/B \to a_0$.  It
will make the ratio $R$ larger than 3 in the four-quark scenario.
Thus this SU(3) symmetry breaking effects in the form factors will
not spoil our method but it will instead improve its applicability.

In heavy meson D/B decays, there is an advantage to apply heavy
quark effective theory. Unlike non-leptonic decays, the SU(3)
breaking effects in semi-leptonic heavy meson decays is quite small,
which is guaranteed by the heavy quark symmetry.  The size of the
SU(3) breaking effect could be roughly estimated by the mass
difference between $u/d$ and $s$ quarks, whose magnitude is
\begin{eqnarray}
 \frac{m_s-m_{u/d}}{\sqrt{m_{D/B}\Lambda}}\sim \left\{\begin{array}{c}
                                                    0.3 {\;\;\rm for \;\; D}\\
                                                    0.1 {\;\;\rm for \;\; B}
                                                  \end{array},\right.
\end{eqnarray}
where $\sqrt{m_{D/H}\Lambda}$ denotes the typical scale in the form
factors, and $\Lambda$ is hadronic scale. Clearly, the $B$ decays
suffer less pollution from the SU(3) symmetry breaking effect. Even
if in the $D$ meson case, the SU(3) breaking effect of 30\% can not
pollute the clear difference between 1 and 3 of ratio $R$ in
eq.(\ref{rule2}).

{ If the mixing angle is close to $\theta=0^\circ$ or
$\theta=90^\circ$ in the two-quark picture  ($\phi=54.7^\circ$ or
$\phi=144.7^\circ$ in four-quark scenario), either $\sigma$ or $f_0$
meson has small production rates but the other one should have large
production rates. Neglecting the highly suppressed channel, the
ratio defined in eq.(\ref{rule1},\ref{rule2}) can still distinguish
the two different scenarios for scalar mesons.}
 If the mixing angle
is modest, i.e. it is not close to the values discussed in the above
paragraph, all three $D^+\to Sl^+\nu$ channels would have similar
branching ratios in magnitude. The branching ratio of the
semileptonic $D_s\to f_0$ decay is measured  as\cite{Ecklund:2009fi}
$ {\cal B}(D_s\to f_0l\bar\nu)\times {\cal B}(f_0\to
 \pi^+\pi^-)
 =(2.0\pm0.3\pm0.1)\times 10^{-3}.
$
Thus as an estimation, branching ratios for the cascade $D^+\to
Sl^+\nu$ decays are expected to have the order
 $\frac{V_{cd}^2}{V_{cs}^2}\times 2\times 10^{-3} \sim 1\times
 10^{-4}$.
The luminosity of BES-III experiment at BEPC II in Beijing is
designed as $3\times 10^{32} {\rm cm}^{-2} {\rm s}^{-1}$. This
experiment, starting running since summer 2008,  will accumulate 30
million $D\bar D$ pair per running year.\cite{Asner:2008nq} Even we
assume the detect efficiency is only $20\%$, there will be 600
events per running year. It is very likely to observe these decay
channels.

As for the $B$ decays, the branching ratio of $B\to Sl\bar\nu$ can
be estimated utilizing the $B\to \rho l\bar\nu$ and $D_s^+\to\phi
l^+\nu$ decays. If the mixing angle is moderate, the branching ratio
can be estimated using heavy quark symmetry as
\begin{eqnarray}
 {\cal B}(B\to f_0l\bar\nu)&\sim& {\cal B}(B\to \rho l\bar\nu)\frac{{\cal B}(D_s\to f_0l\bar\nu)}{{\cal B}(D_s\to\phi
l\bar\nu)}
 ~\sim ~ 10^{-4} \times \frac{10^{-3}}{10^{-2}} =10^{-5}.
\end{eqnarray}
Such a large branching fraction offers a great opportunity for
distinguishing the descriptions. Even if the present $B$ factory
does not observe these channels, it is easy for the forthcoming
Super B factory to measure these channels. Although $\bar B\to
J/\psi S$ are hadronic decays, the hadronic uncertainties are mostly
canceled in the sum rule of ratios.  The branching fraction is
expected to have the order
\begin{eqnarray}
 {\cal B}(B\to f_0J/\psi)&\sim& {\cal B}(\bar B^0\to \rho^0 J/\psi)\frac{{\cal B}(D_s\to f_0l\bar\nu)}{{\cal B}(D_s\to\phi
l\bar\nu)}
 ~\sim ~ 10^{-5} \times \frac{10^{-3}}{10^{-2}}
 =10^{-6}.
\end{eqnarray}
On experimental side, the $J/\psi$ is easily detected through a
lepton pair $l^+l^-$ and thus this mode may be more useful.

%
%


In conclusion, we have investigated the possibility to distinguish
the two-quark and four-quark picture for light scalar mesons. The
semileptonic $D/B\to Sl\bar\nu$ decays and the nonleptonic $B\to
J/\psi(\eta_c) S$ decays are discussed in detail.  These decay
channels have a large potential to be measured on the ongoing
BES-III and the forthcoming Super B experiments. With the same
quantum number as the vacuum, the lightest scalar mesons are very
complicated in nature. It is likely that these scalar mesons are
neither pure 2-quark nor 4-quark states. However, our method is at
least  helpful to rule out one of the possibility. If the ratio $R$
were 1, the pure 4-quark picture is likely to be ruled out but if
the ratio $R$ were 3 the pure 2-quark picture is likely to be
excluded. Our proposed method provides a unique role to uncover the
internal structures of scalar mesons and help to solve the old
puzzles.

\section*{Acknowledgments}

This work is partially supported by National Natural Science
Foundation of China under the Grant No. 10735080, and 10625525. W.
Wang would like to acknowledge Pietro Colangelo for warm hospitality
and useful discussions.


\section*{References}
\begin{thebibliography}{99}


\bibitem{Spanier:2008zz}
  S.~Spanier, N.~A.~Tornqvist and C.~Amsler,
  ``Note on Scalar Mesons,'' Review published on Particle Data
  Group.


\bibitem{Godfrey:1998pd}
  S.~Godfrey and J.~Napolitano,
  Rev.\ Mod.\ Phys.\  {\bf 71}, 1411 (1999).


\bibitem{Close:2002zu}
  F.~E.~Close and N.~A.~Tornqvist,
  J.\ Phys.\ G {\bf 28}, R249 (2002).


\bibitem{Amsler:2008zzb}
  C.~Amsler {\it et al.}  [Particle Data Group],
  Phys.\ Lett.\  B {\bf 667}, 1 (2008).

\bibitem{wl}W. Wang, C.D. Lu, e-Print: arXiv:0910.0613 [hep-ph]
\bibitem{Tornqvist:1995kr}
  N.~A.~Tornqvist,
  Z.\ Phys.\  C {\bf 68}, 647 (1995).


\bibitem{Cheng:2005nb}
  H.~Y.~Cheng, C.~K.~Chua and K.~C.~Yang,
  Phys.\ Rev.\  D {\bf 73}, 014017 (2006); and references therein.



\bibitem{Jaffe:1976ig}
  R.~L.~Jaffe,
  Phys.\ Rev.\  D {\bf 15}, 267 (1977);
  Phys.\ Rev.\  D {\bf 15}, 281 (1977).




\bibitem{Maiani:2004uc}
  L.~Maiani, F.~Piccinini, A.~D.~Polosa and V.~Riquer,
  Phys.\ Rev.\ Lett.\  {\bf 93}, 212002 (2004).



\bibitem{Ecklund:2009fi}
  K.~M.~Ecklund {\it et al.}  [CLEO Collaboration],
  Phys.\ Rev.\  D {\bf 80}, 052009 (2009)
  [arXiv:0907.3201 [hep-ex]].




\bibitem{Asner:2008nq}
  D.~M.~Asner {\it et al.},
  arXiv:0809.1869 [hep-ex].



\end{thebibliography}
\end{document}